# A FUZZY LOGIC-BASED QUALITY MODEL FOR IDENTIFYING MICROSERVICES WITH LOW MAINTAINABILITY

Rahime YILMAZ[1], Feza BUZLUCA[2]

[1] Computer Engineering Department, Istanbul Technical University, Altınbaş University, Istanbul, Turkey yilmazr18@itu.edu.tr
(https://orcid.org/0000-0003-4079-2260)

[2] Computer Engineering Department, Istanbul Technical University, Istanbul, Turkey buzluca@itu.edu.tr
(https://orcid.org/ 0000-0001-9589-8754)

**Abstract**—

Microservice Architecture (MSA) is a popular architectural style that offers many advantages regarding quality attributes, including maintainability and scalability. Developing a system as a set of microservices with expected benefits requires a quality assessment strategy that is established on the measurements of the system's properties. This paper proposes a hierarchical quality model based on fuzzy logic to measure and evaluate the maintainability of MSAs considering ISO/IEC 250xy SQuaRE (System and Software Quality Requirements and Evaluation) standards. Since the qualitative bounds of low-level quality attributes are inherently ambiguous, we use a fuzzification technique to transform crisp values of code metrics into fuzzy levels and apply them as inputs to our quality model. The model generates fuzzy values for the quality sub-characteristics of the maintainability, i.e., modifiability and testability, converted to numerical values through defuzzification. In the last step, using the values of the sub-characteristics, we calculate numerical scores indicating the maintainability level of each microservice in the examined software system. This score was used to assess the quality of the microservices and decide whether they need refactoring. We evaluated our approach by creating a test set with the assistance of three developers, who reviewed and categorized the maintainability levels of the microservices in an open-source project based on their knowledge and experience. They labeled microservices as low, medium, or high, with low indicating the need for refactoring. Our method for identifying low-labeled microservices in the given test set achieved 94% accuracy, 78% precision, and 100% recall. These results indicate that our approach can assist designers in evaluating the maintainability quality of microservices.

**Keywords**—Microservice, microservice quality, quality model, quality measurement, maintainability, fuzzy logic.

## 1. INTRODUCTION

In recent years, the concept of "microservices" has emerged as a widely adopted architectural style in the software industry, praised for productivity benefits [1, 2]. This approach emphasizes deploying small services, called microservices, which shift service-oriented techniques from system integration to system design, development, and deployment [3]. The Microservice Architecture (MSA) facilitates the development of software applications as a collection of microservices that run independently and communicate with each other via lightweight mechanisms using various protocols, such as HTTP, gRPC, or message queue brokers like RabbitMQ and Kafka [4, 5]. This design and development approach offers several benefits, including highly maintainable and scalable software, a smaller codebase, easier software changes, enhanced flexibility for faster delivery, increased team autonomy, and manageable complexity [6].

In MSA, designing maintainable services is essential to obtain its expected benefits. However, maintaining a large number of services can still be challenging. Therefore, it is critical to assess the quality of each service and perform necessary refactoring to ensure that the system is reliable, scalable, interoperable, and maintainable [7, 8]. Despite the diversity of methods to assess software quality with various approaches, there is still a need for comprehensive and applicable quality assessment methodologies for MSA [4, 9]. To address this issue, we propose a hierarchical quality model based on fuzzy logic that uses low-level software code metrics to measure and evaluate the maintainability of microservice architectures. The main objective of our method is to assist developers in identifying microservices with low maintainability levels that require refactoring.

The high-level (external) software characteristics, such as maintainability, can only be assessed after the system is completed and deployed since it is necessary to examine the software's operational dynamics and performance within its operational environment [10, 11]. The effort required to test, modify, and improve a microservice can be determined by testing it in the field, detecting errors, and incorporating new requirements [12]. We built a hierarchical model that connects the maintainability to the related low-level quantifiable properties of the software systems to measure (in fact, to predict) the maintainability of microservices during the development process. The term "low-level properties" refers to the internal properties of a software module or system that can be directly affected by the design and coding decisions made during the development process [13]. Our model follows the ISO/IEC 25010 standard [14] and evaluates microservices' maintainability, focusing on two primary high-level sub-characteristics: modifiability and testability. We determined the metrics to represent low-level properties of a software system that can be used to evaluate the modifiability and testability characteristics of microservices during the development process of the MSA.

While metrics offer valuable insights into the properties of software modules, determining their exact values for use in decision-making models at different levels, e.g., low, medium, and high, can be challenging. We encounter grey areas when interpreting metric values because a value can fall into multiple categories simultaneously. For example, the size of a microservice with 20 methods can be in the grey area between low and medium. In a classical rule-based decision-making system, each crisp value of a metric can only belong to one category, even if it is very close to another level. This can lead to inaccurate evaluation of quality characteristics, especially those with a value at the border. To solve this problem, we build a fuzzy logic system to model logical reasoning in the context of imprecise values where a metric's value can belong to multiple categories simultaneously [15, 16]. Additionally, fuzzification allows us to convert crisp metric values to linguistic variables and levels such as "low coupling" and "high complexity". Using linguistic variables, we construct inference rules in natural language based on human reasoning to evaluate modifiability and testability levels of microservices, e.g., "If the internal coupling is high and complexity of a microservice is high, then its modifiability is low." In the last step of our measurement system, we apply a defuzzification technique to convert the results of the inference rules into crisp numerical values and combine them to calculate the maintainability score for each microservice in the system under evaluation. Thus, this score is used to assess the quality of the microservices and decide if they need refactoring.

To evaluate our approach, we created a test set of microservices with the help of three developers who evaluated and categorized 36 microservices in the open-source "Train Ticket" project [17] based on their experience and knowledge. Our model identified all low-labeled (LOW) microservices that require refactoring, achieving a 100 % recall score. Since two medium-labeled (MED) microservices have also been identified as LOW, the method's precision is 78%. We also observed that the obtained scores were consistent with the labels MED and high-labeled (HIGH). The results indicate that the proposed method can be utilized to assess the quality of microservices in terms of their maintainability. It can effectively enhance microservice-based applications by identifying the causes of low-level maintainability and providing feedback to the development team.

The summary of our contributions in this study is as follows:
- A hierarchical model was built that connects the sub-characteristics of maintainability, i.e., modifiability and testability of microservices, to the related low-level static software metrics that can be used to measure (predict) the maintainability during the development process.
- A measurement system based on fuzzy logic was built to handle the imprecise values of metrics that can belong to multiple categories simultaneously. We constructed a fuzzification process to convert crisp metric values to linguistic levels and defined inference rules in natural language based on human reasoning to evaluate the modifiability and testability levels of microservices.
- Our approach was evaluated on an open-source MSA-based project by comparing the results of our method with the opinions of three human evaluators.

The rest of the paper is organized as follows: Section 2 explores the current literature and relevant studies. Section 3 provides an overview of the hierarchical quality model and the implementation of fuzzy logic systems. Section 4 outlines the proposed hierarchical model, explaining in detail the procedural aspects of the fuzzy logic system and the method used for evaluating maintainability. Section 5 presents the experimental results obtained through the application of the proposed quality model, together with the validation process for these results. Section 6 comprehensively discusses the findings, while Section 7 describes the potential limits and the general validity of our research. Section 8 concludes the study by providing last remarks and proposing directions for future research.

## 2. RELATED WORK

Microservices' maintainability is a topic of interest due to its significance, and there are research studies available on the quality of Microservices Architecture (MSA). An early metric-based quality model for evaluating the maintainability of services- and microservices-based systems was proposed by Bogner et al. [18]. In their pioneering study, they defined microservice maintainability as a collection of service properties measured by service metrics. The authors' study highlights the necessity for practical Quality Models in Service-Based and Microservice-Based Systems, proposing a hierarchical model encompassing top-level attributes for maintainability, intermediate layers for service properties (SPs), and lower levels for service metrics associated with specific SPs. They introduced a hierarchical model but had not undergone a complete evaluation at the time of publication. Diverging from the study in [18], our quality model is built as a fuzzy logic system for measuring and evaluating the maintainability of individual microservices. In addition, our study involves experimentation on projects sourced from GitHub to evaluate the performance of the proposed measurement system. Following the initial study, Bogner et al. enhanced and extended their research. Their next publication addresses the challenge of automatically calculating service-based maintainability metrics by analyzing the runtime data in microservice-based systems [19]. The demonstrated approach in [19] proves successful in a small case study, but careful evaluation in larger systems and consideration of combining dynamic and static information are suggested for future work. Another study published by Bogner et al. aims to calculate maintainability metrics based on machine-readable interface descriptions of RESTful services [20]. They introduced a designed approach (called RAMA) to derive actionable thresholds for metrics, aiming to enhance the application of metrics implemented within the approach. In that study [20], the authors utilized quartile distribution to define labels for their metrics, creating four ranked bands based on the distribution of metric values. Similarly, in our study, we also employed median quartiles for fuzzification parameters to determine the fuzzification parameters. In contrast, our approach allows a metric value to belong to multiple categories at the same time.

In several proposals, researchers have used metrics to measure quality characteristics directly, employing quality models like QMOOD. Apel et al. [21] developed a methodology to evaluate software quality in microservice architectures by deriving a set of metrics from MOOD [22] and QMOOD [23], performing specific operations in combination with their considerations. This study [21] focuses on the impact of metrics on key software quality characteristics, such as maintainability, performance efficiency, functional suitability, and reliability defined by ISO/IEC 25010 for the consideration of microservice architecture. The authors utilize their case study to investigate which metrics are suitable for evaluating these characteristics and how variations in metric values impact overall software quality. In their model, the relationship between the metrics and software quality characteristics is established through qualitative data evaluation and prior discussions rather than using MOOD or QMOOD reference evaluations. As in our study, Apel et al. established a correlation between quality characteristics defined in ISO standards and metrics, although their study did not include an evaluation system. Their findings indicated which metrics contributed to maintainability and to what extent, whereas our proposal is a measurement system based on a relationship between low-level metrics and the maintenance score of each microservice, as well as an evaluation system based on these scores.

Cardelli et al. [7] proposed MicroQuality, a novel approach for the quality evaluation of MSAs. This approach in [7] outlines the specification, aggregation, and evaluation of software quality attributes to evaluate microservice-based systems. Employing the MicroQuality quality evaluation framework, the system's maintainability was measured by analyzing the coupling, cohesion, and complexity of its microservices through the utilization of a benchmark system. The approach assesses the maintainability of microservices-based systems by employing model-driven engineering techniques; however, detailed measures are not explicitly provided. While their approach focuses on analyzing single elements in the MSA model and evaluates the maintainability of the overall system, our approach conducts a detailed evaluation for each microservice independently, offering a comprehensive assessment at the service level.

Subsequently, Milic et al. [24] explore monolithic and microservice software architectures, evaluating the software system using Feature-driven development. They defined the formulation of a quality-based mathematical model for software architecture optimization, utilizing a continuous quality assessment approach that evaluated these architectures. This study examines coupling, testability, security, complexity, deployability, and availability quality attributes. The authors proposed a set of quality metrics for monolithic and microservice architectures, considering their alignment with system architecture, to facilitate a well-informed selection of the most suitable architecture for a particular software system. Their model does not provide detailed measurements, but it identifies key attributes similarly to ours, based on quality characteristics and sub-characteristics to specify software quality attributes.

In another study, Pulnil et al. [4] introduced a microservices quality model, employing the QMOOD method and focusing primarily on microservice antipatterns. This paper builds upon their work by extending the QMOOD method to incorporate microservices antipatterns along with the ISO/IEC 25010 standards for quality attributes to present a microservices quality model. Both that study and our study adopt a hierarchical structure for evaluations and utilize the Train Ticket microservices project for validation by comparing the design quality of experimental systems. While Pulnil et al. evaluate multiple characteristics such as maintainability, scalability, security, and reliability, our study specifically concentrates on maintainability. Pulnil et al. calculate quality characteristic scores before and after refactoring to gauge the success of the refactoring process, while our model focuses on assessing maintainability scores to identify microservices in need of refactoring. Our model evaluates each microservice's maintainability using our proposed quality model, whereas Pulnil et al. assess the overall improvement of the MSA after refactoring the projects and provide maintainability scores for the whole project pre- and post-refactoring.

A recent study published by Hasan et al. [25] suggests using metrics derived from cloud-native design principles to evaluate how software can be maintained when transitioning from monoliths to microservices. This evaluation was based on various design properties like coupling, cohesion, complexity, and size considering ISO/IEC 25010 standard. The study provides designers and developers with a way to assess the maintainability of microservices during the design phase, and it incorporates various design properties that are developed by recent research in this area. Despite the study presenting a service-based evaluation model for maintainability, the results of the study were not obtained and have been left to be studied in the future. In addition, the proposed model does not contain sub-characteristics, contrary to ours.

Lastly, in our previous work, we proposed a preliminary hierarchical quality model to assess microservice maintainability [26]. The main enhancement of the new study lies in the field of fuzzy logic-based measurement. The previous model contains a limited fuzzy logic system with a basic fuzzification process, inference rules, and no defuzzification. As the fuzzification process in [26] uses only the singleton membership function, it behaves like the standard logic, and each metric value can only be included in one of the sets LOW, MED, or HIGH. However, in the new study, we improved the fuzzification process by defining triangular and trapezoidal functions to quantify quality properties. This leads to a metric value having membership degrees in multiple fuzzy sets at the same time, allowing for the management of ambiguous areas within the metric values. Since the method in [26] does not contain a defuzzification stage, the maintainability level of microservices is defined only using linguistic levels. In this study, a defuzzification process is designed to assign crisp values for the quality measures of sub-characteristics modifiability and testability that are used to calculate the maintainability score. Thus, the new approach enables a more precise evaluation of the maintainability of the microservices.

The study in [26] focuses on the sub-characteristics of testability and modularity to measure maintainability. However, our new approach considers modifiability instead of modularity because modifiability is strongly related to other sub-characteristics of maintainability, such as modularity, reusability, and analyzability [14]. For example, a modular software system is also modifiable. Similarly, an easily modifiable software module is likewise easily analyzable. Accordingly, the metrics used in the new method have also changed.

## 3. BACKGROUND

This section covers the background of our study. We will explore key elements related to hierarchical quality models and the application of fuzzy logic. Both of these subjects play a crucial role in influencing our methodology for accurately evaluating the maintainability of microservices.

### 3.1. Hierarchical Quality Models

Usually, we need to evaluate the high-level characteristics of software systems that represent the needs of various stakeholders, such as maintainability. However, direct measurement of the high-level, external software characteristics is only possible after the system is completed and deployed [10]. The realistic effort required to maintain a microservice can be determined when it is modified, tested, and improved in the field [12]. To measure (in fact, to predict) the high-level characteristics during the development process, hierarchical models are built to determine the related low-level properties of the software systems that affect the high-level characteristics under discussion. The low-level properties, such as the size of a software module, depend on the design and coding of the system, and they can be directly measured during the development process [13].

ISO/IEC 25010 [14] defines a set of terms describing software quality and provides a hierarchical model that categorizes software product quality into characteristics that are further subdivided into sub-characteristics. Maintainability is one of the main characteristics in the product quality model and it contains five sub-characteristics: modularity, reusability, analyzability, modifiability, and testability. Our model follows ISO/IEC 25010 and evaluates microservices' maintainability, focusing on two primary sub-characteristics: modifiability and testability.

Our measurement approach is built based on the Quality Measurement Reference Model (QM-RM) introduced in ISO/IEC 25020 [27]. This model describes how to evaluate high-level quality characteristics depending on the low-level, measurable properties of a system. The measurable quality properties (QP) of a system/software product are called properties to quantify and can be associated with quality measures. These properties are measured by applying a measurement method that is a logical sequence of operations used to quantify properties according to a specified scale. The value obtained from a measurement method is called a quality measure element (QME). QMEs are combined appropriately using a measurement function, and as a result, quality measures (QMs) are obtained. QMs represent the quantifications of the quality sub-characteristics and characteristics we want to evaluate.

The ISO standards provide a shared language among professionals in the field, enhancing clarity and understanding. However, they only describe the measurement model and do not provide related software properties to quantify. Therefore, we defined the related static software metrics and their connections to sub-characteristics modifiability and testability. The proposed quality model and measurement procedures are detailed in Section 4.

### 3.2. Fuzzy Logic

Fuzzy logic is a subfield of mathematical logic that provides a framework for dealing with situations with no clear boundary between true and false or where there are multiple levels of truth [28]. It operates on the premise of employing linguistic variables associated with a fuzzy set. Fuzzy sets represent uncertain concepts such as "low, medium, and high," which cannot be expressed through standard sets and crisp values [16]. We build a fuzzy logic system that consists of three main parts, i.e., fuzzification, inference based on rules, and defuzzification, as presented in Figure 1.

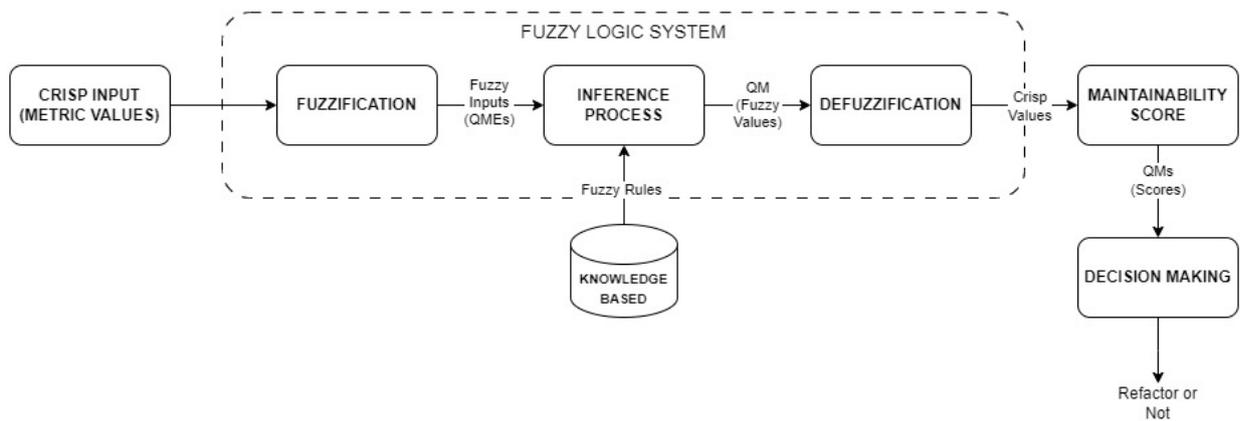

**Fig 1** General structure of a fuzzy logic system

*Fuzzification:*

Due to the intangible nature of software quality, it is hard to establish exact threshold values for metrics such as lines of code, number of methods, coupling, etc. Therefore, when building a hierarchical quality model, we can classify the quality attributes they represent as ordinary or exceptional. In the case of a metric representing coupling among software classes in a microservice, if the value is 25, would we classify it as low, normal, or high? What is the maximum number of methods per microservice beyond which it becomes too large? There may be grey areas when interpreting metric values since a value may belong to more than one category simultaneously. For example, the size of a microservice with 150 lines of code can be perceived as normal by some developers, while others may interpret it as excessive. In a classical rule-based decision-making system, each metric value can only belong to one category, such as LOW, MED, or HIGH, even if it is very close to another category. This can result in incorrect assessment of quality characteristics, particularly those with a value at the boundary. On the other hand, the fuzzification process, which is the first part of a fuzzy system, allows for values to have degrees of membership to a fuzzy set rather than being strictly true or false. Fuzzy sets represent uncertain concepts such as "low, medium, and high", and a particular metric value can belong to multiple fuzzy sets with different degrees simultaneously [16]. To determine the membership degrees of a crisp value, membership functions are defined for each fuzzy set. Each membership function generates an output value between 0 and 1, indicating the strength of the membership to the corresponding fuzzy set. For example, the membership function $\mu_A(x)$ will provide the membership degree of the value x to set A.

*Fuzzy Inference Rules:*

The second part of a fuzzy logic system consists of the Fuzzy Inference Module (FIM), where IF-THEN rules are applied to the membership degrees of inputs to obtain fuzzy results for outputs. The rules in an FIM are typically in the form of "IF-THEN" statements, where IF (antecedent part) represents the fuzzy input, and THEN (consequent part) represents the fuzzy output [29]. Fuzzy inference rules are usually determined by the combination of expert knowledge and empirical data, as well as by carefully considering the linguistic terms used to describe the variables and their respective weights. Inference combines the membership degrees generated by fuzzy rules to yield an overall output or conclusion [30]. Since the fuzzification process converts crisp input values to linguistic variables and levels such as "coupling is low" and "complexity is high", the inference rules can be defined in natural language based on human reasoning. For example, in our study, we defined an inference rule as follows: "If the internal coupling is high and the complexity of a microservice is high, then its modifiability is low."

***Defuzzification:***

The last part of a fuzzy system is the defuzzification that converts the outputs of the FIM into crisp numerical values. In our study, we applied the centroid defuzzification technique[31] to combine and convert the membership degrees obtained from the FIM into crisp numerical values, indicating the modifiability and testability levels of each microservice.

## 4. THE PROPOSED METHOD

In our study, we first constructed a hierarchical quality model that connects various low-level software-oriented properties of microservices to their corresponding quality characteristics, testability, and modifiability that determine the maintainability. Then, we built a fuzzy logic system to measure and evaluate the maintainability of microservices based on the values of code metrics that represent low-level quality properties. In the following sub-sections, 1, 2, and 3, we explain the proposed quality model, the fuzzy logic system, and the calculation of maintainability, respectively.

### 4.1. The Proposed Hierarchical Quality Model

Figure 2 illustrates the general structure of the proposed hierarchical model that links the microservices' external (high-level) quality characteristic maintainability to their internal (low-level) properties. The model is constructed in a top-down fashion. First, we determined the major sub-characteristics of the maintainability to focus on, i.e., modifiability and testability. Secondly, we identified the low-level properties of microservices that affect their sub-characteristics and depend on their design and coding. In the last step, we defined the software metrics that can be quantified based on the selected properties of microservices. The details of these steps are explained in the following sub-sections 4.1.1, 4.1.2, and 4.1.3.

We built a measurement system based on fuzzy logic to establish the quantitative connections between the layers of the hierarchical model. The measurement system operates in a bottom-up fashion. In the first step, a fuzzification process is carried out on the crisp metrics values to obtain quality measurement elements (QMEs) for each property as membership degrees to fuzzy sets in three categories: LOW, MED, and HIGH. Secondly, using a measurement function based on the proposed inference rules and defuzzification technique, QMEs are combined to obtain quality measures (QMs) as quantitative measurements of the sub-characteristics modifiability (QM1) and testability (QM2). In the final step, we calculate the weighted average of these sub-characteristics to get the QM of the main quality characteristic maintainability for each microservice. This QM is used to decide whether a microservice requires improvement. The details of the fuzzy logic system we built for measurement are explained in sub-section 2.

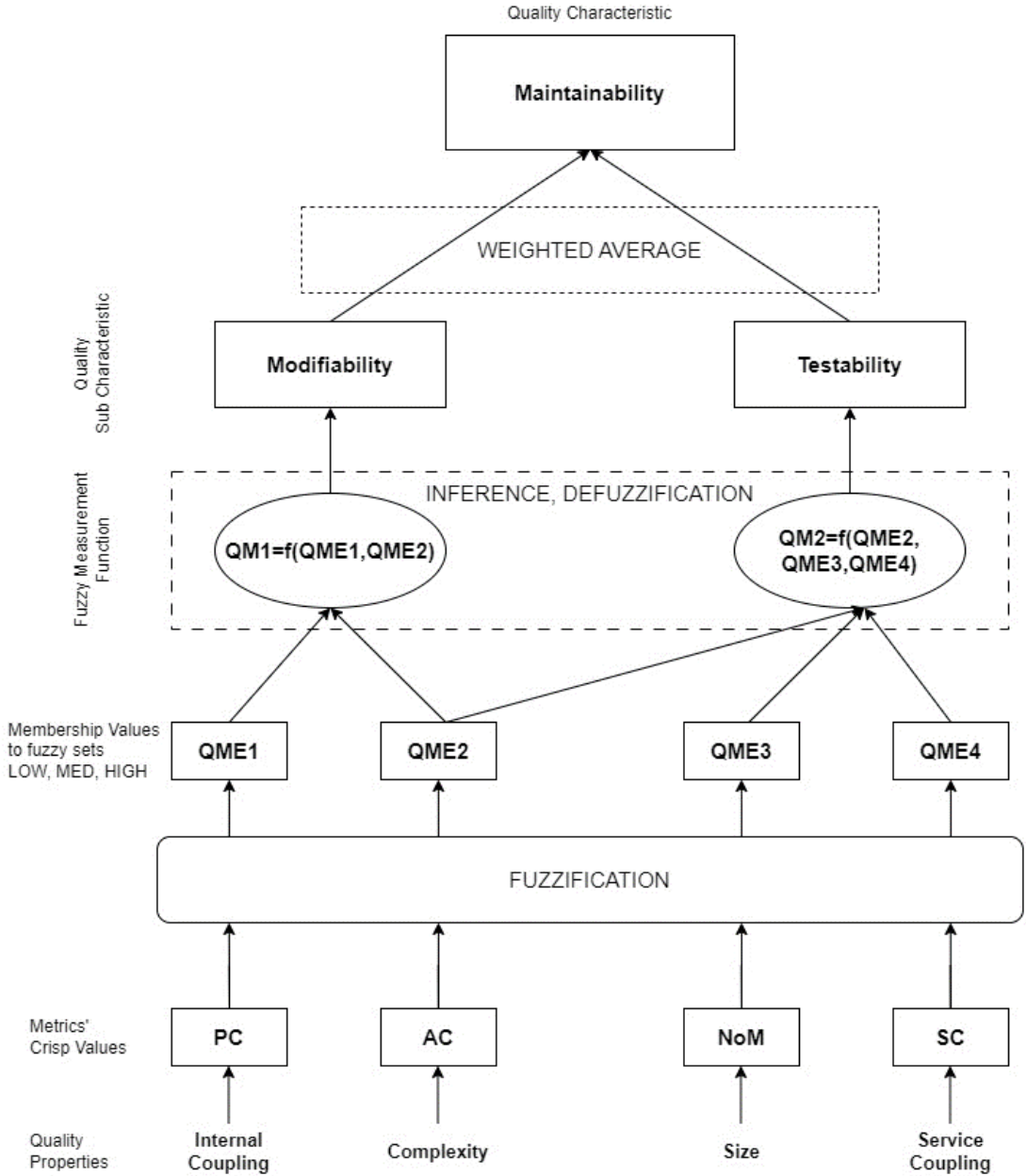

**Fig 2.** Proposed hierarchical quality model for MSA

### 4.1.1. Quality Characteristics for Microservice Architecture

The maintainability quality attribute usually manifests many months after the initial development has been finished, and costs for performing software maintenance activities take a very large fraction of the complete development budget [32, 33]. Therefore, this study focuses on maintainability, one of the most important quality attributes to guarantee the efficient handling of long-living software systems, including microservice-based architectures [34-36].

The ISO/IEC 25010 defines five sub-characteristics for maintainability: modularity, reusability, analyzability, modifiability, and testability. In this study, we consider two major sub-characteristics, modifiability and testability, that are essential for the maintainability of microservices. These two characteristics are also indicators of other attributes, i.e., a microservice that can be modified and tested is likely to be modular, analyzable, and reusable [37]. However, the proposed approach can also be employed to construct a more detailed model containing other sub-characteristics. In the following paragraphs, we explain the meaning and significance of modifiability and testability for the maintenance of microservices.

*Modifiability:*

In microservices, modifiability is a significant quality attribute because it allows the system to evolve over time to accommodate new features, adapts to changing business needs, and fix bugs without requiring a complete overhaul of the entire system [38]. Modifiability is also an indicator of other quality sub-characteristics such as modularity, reusability, and analyzability as a modifiable microservice typically meets these criteria, too [14]. To be modifiable, a microservice should have an understandable, flexible internal design with minimal dependencies [11]. This makes it easier to make changes to a microservice without introducing defects or degrading existing product quality.

*Testability:*

Testability is the characteristic that measures the degree to which a system, component, or unit is able to be tested [14]. In the context of microservices, testability includes the ability to test individual microservices independently from the rest of the system and in the context of their interactions with other services. It is crucial to ensure that the microservices can be easily tested to improve the system's overall quality and reliability [39].

### 4.1.2. Quality Properties for Microservice Architecture

The low-level internal quality properties are fundamental dimensions for assessing and evaluating the quality of software systems. Many properties affect a quality characteristic to varying degrees. However, using too many features makes the model incomprehensible and does not significantly increase the measurement accuracy. In this study, we examined four key quality aspects of microservices that significantly influence their ability of modifiability and testability [40, 41]. These quality properties are internal coupling, service coupling, size, and complexity. Modifiability is evaluated based on internal coupling and complexity, while testability is evaluated using service coupling and complexity. Below is a detailed explanation of the quality properties and their relations to high-level quality characteristics.

*Internal Coupling:*

Internal coupling refers to the degree to which different components, e.g., classes within a single microservice, depend on or interact with one another. Since a microservice is composed of related and cooperating classes, it is normal for internal coupling to be high to a certain degree. It is also considered to be an indicator of a microservice's cohesion. However, if dependency among classes gets too high, the required effort to modify a microservice also increases [42]. Therefore, a high degree of internal coupling can reduce the modifiability of microservices in our model.

*Service Coupling:*

Service coupling refers to the degree to which different services depend on or interact with one another. When one service needs to utilize the functionality of another service, it makes a service call to that service, passing any necessary information or data as part of the request. The service that receives a call processes the request and returns a response to the calling service. In this scenario, there is a dependency between two services, and the calling service is coupled to the receiving service. One of the goals of designing a microservices architecture is to create loosely coupled services to make them more independent and easier to test and scale [43]. If the service coupling of a microservice is high, it is difficult to understand and test it independently from other services. Service coupling was selected as a factor in our model to evaluate the testability and, consequently, the maintainability of microservices.

*Complexity:*

Chidamber and Kemerer defined class complexity as the Weighted Methods per Class (WMC) metric [22]. Various variants of the WMC metric, such as Average Method Complexity (AMC), have also been used to measure complexity [44]. In this study, we adopted the AMC metric for microservices and defined the complexity of a microservice as the average complexity of its methods (functions). In software development, complexity is considered one of the most significant quality properties since high complexity adversely impacts many quality characteristics, such as modifiability, testability, and, consequently, maintainability [45].

*Size:*

Microservice size plays a crucial role in designing an MSA as it directly impacts the expected benefits [46]. One significant factor influencing microservice size is the number of functions it encompasses. Generally, an increase in the number of functions within a microservice leads to a larger microservice size due to the addition of code and logic [47]. A microservice with too few functions may result in underutilization of its capabilities, while a microservice with too many functions may demand more testing effort. Testing a microservice with an excessive number of methods can present difficulties, potentially resulting in incomplete test coverage [8]. In addition, such a process can be resource-intensive and time-consuming. Determining the optimal size for microservices involves finding the right balance between the number of functions it can perform and the ease with which it can be tested. The right balance allows for enhanced development, testing, and deployment processes, leading to more efficient microservices [48]. Thus, having an appropriate number of methods is necessary as a result of the responsibilities and functionalities of the microservice.

### 4.1.3. Microservice-based Metrics

To quantify the quality properties of microservices, we analyzed various published service-based metrics and selected the Propagation Cost (PC) for internal coupling, Service Call (SC) for service coupling, Average Complexity (AC) for complexity, and Number of Methods (NoM) for size. The service call metric was obtained by analyzing the source code, while the other metrics were obtained using the SonarGraph and JArchitect tools. Detailed explanations of these metrics are as follows:

Propagation Cost (PC): It refers to the incurred overhead associated with the propagation or communication of changes from one element to other elements in the system that depend on it [49]. In our study, this metric provides an initial insight into the degree of coupling within a microservice, with higher values indicating a higher degree of coupling and, consequently, potentially higher propagation costs. Propagation cost varies depending on the number of classes in a microservice and the dependencies and interactions between them. It is essentially the average component dependency divided by the total number of elements in the microservice.

Service Call Ratio (SC): A service call for microservices refers to a request made by one microservice to another microservice to access or utilize its functionality[50]. To calculate the number of service calls a specific microservice makes, we count the instances where that microservice acts as the service caller, making requests to other microservices. The number of microservices in a project varies from project to project; thus, the relationships between services may vary at the same rate. To calculate this metric, we divide the number of calls made by a microservice by the total number of microservices in the project, making it independent of project size.

Average Complexity (AC): It refers to the cumulative complexity of microservices. The average complexity metric in SonarGraph is calculated by summing up the cognitive complexity values of all code elements (such as classes, methods, and functions) within the microservice and dividing it by the total number of code elements [51]. Cognitive complexity is calculated based on the control flow and nesting of the code block of microservices.

Number of Methods (NoM): It refers to the number of distinct functions or operations implemented within a microservice. It is important to strike a balance when determining the number of methods within a microservice to provide both functionality and testing. Since the method number is considered to be strongly correlated with the size of the classes within the microservice, NOM is used as an indicator of the size of the microservice [52]. Testing each method in a microservice with many methods will increase the testing effort.

The metrics AC and NoM are barely enough to completely comprehend, considering the significance of contextual elements such as granularity and the distinct function of each microservice. We initially chose NoM and AC due to their objective nature and their proven utility in traditional software engineering research as indicators of code complexity and maintainability[51]. Additional or alternative metrics can be included to construct a model that also captures the granularity and types of microservices. However, whereas the relevance of microservice granularity is widely acknowledged, a commonly accepted definition of this attribute is still lacking [53].

These metrics provide a quantitative basis for evaluating the quality properties of microservices. The values of the metrics are collected from the system under evaluation and given as inputs to the measurement system.

### 4.2. The Structure of the Fuzzy Logic System

The fuzzy logic system was developed to measure the quality characteristics of microservices based on their low-level quality properties and consists of three steps, i.e., fuzzification, inference mechanism, and defuzzification. A detailed description of each of these steps is provided in the subsequent subsections.

#### 4.2.1. Fuzzification of Metrics

In our approach, fuzzification provides QME values of metrics as membership degrees to fuzzy sets LOW, MED, and HIGH. Since we have three fuzzy values, we defined three membership functions: $\mu_L^i(x)$, $\mu_M^i(x)$, and $\mu_H^i(x)$ for each metric (i = PC, AC, NoM, SC) to calculate its membership degrees to related fuzzy sets. Here, the subscript denotes the fuzzy set, i.e., L: LOW, M: MED, and H: HIGH. The superscript is the abbreviation of the related metric, i.e., PC, AC, NoM, and SC. For example, the function $\mu_L^{AC}(x)$ is used to calculate the membership degree of the metric AC to the fuzzy set LOW for a given value x.

QME value for each metric $i$ of a microservice is a triple, which contains the membership degrees calculated during fuzzification, i.e., $QME^i = \{\mu_L^i(x), \mu_M^i(x), \mu_H^i(x)\}$, where $i$ = 1:PC, 2:AC, 3:NoM, 4:SC. For example, if the metric AC of a microservice has the value x=3.8, the functions generate the outputs $\mu_L^{AC}(x) = 0.5$, $\mu_M^{AC}(x) = 0.5$, and $\mu_H^{AC}(x) = 0$, then QME2 = {0.5, 0.5, 0}. The sum of the membership degrees for a given value $x$ of a metric is one as presented by the following equation:

$$\mu_L^i(x) + \mu_M^i(x) + \mu_H^i(x) = 1, \ \forall\ i \in \{PC, AC, NoM, SC\} \qquad (1)$$

To define the membership functions, we selected two common forms that are commonly used to build fuzzy systems [54]. We defined trapezoidal membership functions for the LOW and HIGH sets and triangular membership functions for the MED set, as explained below.

The trapezoidal membership functions $\mu_L^i(x)$ for the fuzzy set "LOW":

The trapezoidal membership functions of the metrics are defined by four points (parameters), i.e., L1$^i$, L2$^i$, L3$^i$, and L4$^i$, which determine its shape. In our model, the lowest two points, L1$^i$ and L2$^i$, of the functions $\mu_L^i(x)$ for all metrics (i = PC, AC, NoM, SC) are zero. The parameters L3$^i$ and L4$^i$ are determined for each metric $i$ through statistical analysis of reference projects. The equation for $\mu_L^i(x)$ is given below and its shape is presented in Figure 3.

$$\mu_L^i(x) = \begin{cases} 1, & L1^i = L2^i = 0 \leq x \leq L3^i \\ \dfrac{(L4^i - x)}{(L4^i - L3^i)}, & L3^i < x \leq L4^i \\ 0, & x > L4^i \end{cases} \quad (2)$$

The triangular membership functions $\mu_M^i(x)$ for the fuzzy set "MED":

The shape of triangular membership functions is determined by three points (parameters), i.e., M1$^i$, M2$^i$, and M3$^i$, that are calculated for each metric $i$ based on statical analysis on reference projects. The equation for $\mu_M^i(x)$ is given below, and its shape is presented in Figure 3.

$$\mu_M^i(x) = \begin{cases} \dfrac{(x - M1^i)}{(M2^i - M1^i)}, & M1^i \leq x \leq M2^i \\ \dfrac{(M3^i - x)}{(M3^i - M2^i)}, & M2^i < x \leq M3^i \\ 0, & x > M3^i \end{cases} \quad (3)$$

The membership functions $\mu_H^i(x)$ for the fuzzy set "HIGH":

The trapezoidal membership functions are defined by four points, i.e., H1$^i$, H2$^i$, H3$^i$, and H4$^i$, for each metric (i = PC, AC, NoM, SC). In our model, the two highest points, H3$^i$ and H4$^i$, of the functions $\mu_H^i(x)$ are set to the highest value MAX$^i$ of the corresponding metric $i$, i.e., H3$^i$ = H4$^i$ = MAX$^i$. The parameters H1$^i$ and H2$^i$ are determined based on statistical calculations on data obtained from reference projects. The equation for $\mu_H^i(x)$ is given below and its shape is presented in Figure 3.

$$\mu_H^i(x) = \begin{cases} 0, & H1^i \leq x \\ \dfrac{(x - H1^i)}{(H2^i - H1^i)}, & H1^i < x \leq H2^i \\ 1, & H2^i < x \leq H3^i = H4^i \end{cases} \quad (4)$$

Since the membership degrees for a given value $x$ of a metric is one, as presented in Equation 1, some fuzzification parameters overlap as follows: $L3^i = M1^i$, $L4^i = M2^i = H1^i$, and $M3^i = H2^i$. Figure 3 illustrates the general shapes of membership functions.

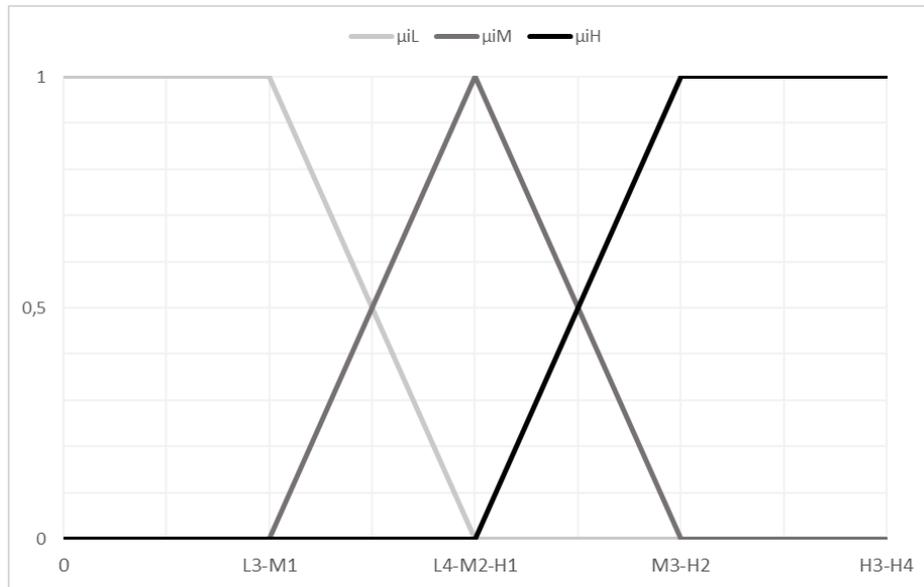

**Fig 3.** General shapes of the membership functions

Table 1 presents linguistic variables, the types of member functions, and their parameters.

| Linguistic Variable | Membership Function | Points, Intervals |
|---|---|---|
| LOW | Trapezoidal | [L1=L2=0, L3, L4] |
| MED | Triangular | [M1, M2, M3] |
| HIGH | Trapezoidal | [H1, H2, H3=H4=MAX] |

**Table 1.** Elements of the fuzzification process

*Determining The Parameters of The Membership Functions:*

Determining parameters for membership functions in the fuzzification process is crucial for accurately representing uncertainty and optimizing the performance of fuzzy logic systems [55]. In our study, we attempt to determine LOW, MED, and HIGH values of metrics that represent the general characteristics of microservices in MSA projects. To determine intervals of the membership functions, $\mu_L^i(x)$, $\mu_M^i(x)$, and $\mu_H^i(x)$, we analyzed metric values of 14 microservices obtained from two MSA-based open-source reference projects developed in Java, namely "TeaStore"[1] and "Microservice Observability"[2]. "TeaStore" is a microservice-based test and reference application that contains six microservices. The "Microservice Observability" project was developed to demonstrate observability patterns for microservices architecture and contains eight microservices. For the evaluation system to function correctly, microservices in reference projects must have varying levels of quality characteristics. Thus, the microservices selected from reference projects were intentionally diverse, representing a spectrum of quality characteristics. The potential bias in selecting projects based on observed quality characteristics may impact the validity of the measurements. The proposed model's parameters may lack accuracy if the reference projects predominantly represent either high or low-quality microservices. In addition, when establishing a measurement system to evaluate software systems for specific applications containing microservices with specified design elements, it is advisable to choose reference projects with similar features to gather the general characteristics of these applications. To clarify, design elements refer to the architectural patterns, principles, and practices used in designing and implementing microservices; similar features pertain to the general structure (functionalities, capabilities, communication mechanisms, etc.) of the reference projects and specific development technologies being used for comparison, and general characteristics denote the typical or common attributes and qualities associated with the microservices in the reference projects.

---
1 https://github.com/DescartesResearch/TeaStore
2. https://github.com/aelkz/microservices-observability/tree/master

Our analysis revealed that the PC, SC, and NOM metrics followed a typical statistical pattern known as a normal distribution. To obtain the fuzzification parameters, we computed three distinct medians for each metric, namely the first quartile (Q1), median (Q2), and third quartile (Q3). A metric array's first quartile (Q1) corresponds to the L3 and M1 parameters for that metric since we consider values less than Q1 as LOW. As the metric values around Q2 are considered to be normal, the median (Q2) corresponds to L4, M2, and H1 parameters. The third quartile (Q3) corresponds to M3 and H2 because values greater than Q3 are considered high. L1 and L2 are always zero, and H3 and H4 are set to the highest value of the related metric. During our manual investigation of the distribution of the AC metric, it was noticed that certain values tend to cluster together. The analysis of the code reveals that there are groups of microservices with LOW, MED, and HIGH complexity. Based on the metric values of the microservices in these groups, we determined well-rounded and comprehensive boundaries for the AC metric. As a result of our analysis of reference projects, we obtain the parameters of the membership functions for the metrics as presented in Table 2.

|     | L1=L2 | L3=M1 | L4=M2=H1 | M3=H2 | H3=H4 |
|-----|-------|-------|----------|-------|-------|
| SC  | 0     | 0.125 | 0.25     | 0.5   | 1     |
| PC  | 0     | 19.4  | 24.2     | 27.1  | 50    |
| AC  | 0     | 2.81  | 4.78     | 5.63  | 11    |
| NoM | 0     | 9     | 16       | 30    | 90    |

Table 2. Parameters of membership functions for each metric

As an example, the membership functions of the Propagation Cost (PC) metric are presented in Figure 4. For example, if a microservice in the system under evaluation (SUE) has a PC value smaller than 19.4, the propagation cost of this microservice is definitely LOW, i.e., $\mu_L^{PC}(19.4) = 1$, $\mu_M^{PC}(19.4) = 0$, and $\mu_H^{PC}(19.4) = 0$. The propagation cost of a microservice with a PC value of 21.8 is between LOW and MED because $\mu_L^{PC}(21.8) = 0.5$, $\mu_M^{PC}(21.8) = 0.5$, and $\mu_H^{PC}(21.8) = 0$.

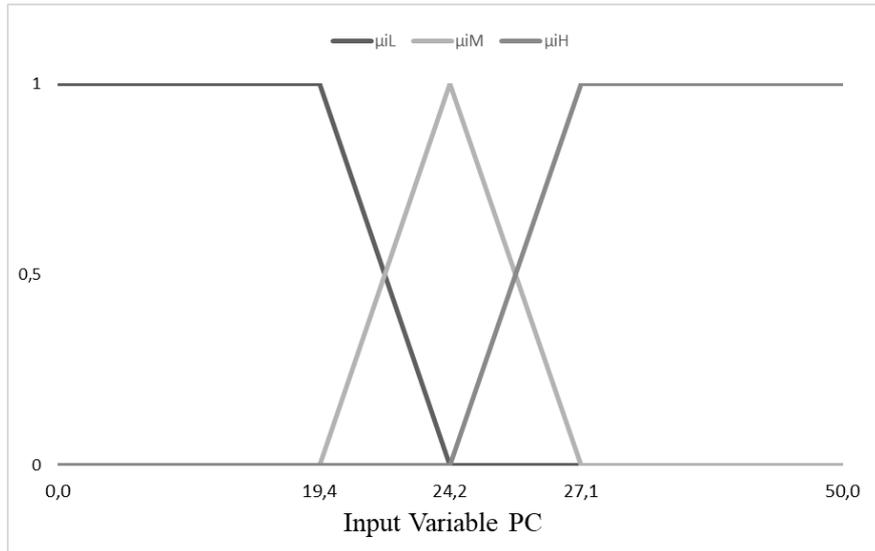

**Fig 4.** Membership functions of the Propagation Cost (PC) metric

### 4.2.2. Fuzzy Inference Mechanism

We generated the inference rules by combining the knowledge obtained from the literature [56, 57], our experience, and commonly accepted perceptions about the relationship between software metrics and quality attributes. Table 3.a presents 9 fuzzy inference rules (RM1 - RM9) used to determine the modifiability level, and Table 3.b contains 27 rules (RT1 – RT27) that determine the testability levels of microservices depending on the fuzzy values of their internal properties. These rules express relationships considering understandable linguistic terms. For example, according to rule RM3 in Table 3.a, if the internal coupling and complexity of a microservice are "HIGH", then its modifiability is "LOW". This rule represents the situation where the classes in a microservice are tightly coupled, and their methods are complex. In this case, the effort required to modify the associated microservice will be excessive, and as a result, its modifiability level is "LOW". According to the rule RT22 in Table 3.b, if the complexity, size, and coupling of a microservice are all "LOW", then its testability is "HIGH". It applies if methods in a microservice are simple, its size is relatively small, and it is not strongly coupled to other services. Since it will be easy to test such a microservice, its testability level is considered "HIGH".

| Rule | Complexity (AC) | Size (NoM) | Service Coupling (SC) | Testability |
|------|-----------------|------------|------------------------|-------------|
| RT1  | MED  | HIGH | LOW  | LOW |
| RT2  | MED  | HIGH | MED  | LOW |
| RT3  | MED  | HIGH | HIGH | LOW |
| RT4  | HIGH | LOW  | HIGH | LOW |
| RT5  | HIGH | MED  | LOW  | LOW |
| RT6  | HIGH | MED  | MED  | LOW |
| RT7  | HIGH | MED  | HIGH | LOW |
| RT8  | HIGH | HIGH | LOW  | LOW |
| RT9  | HIGH | HIGH | MED  | LOW |

**a.** Modifiability

| Rule | Complexity (AC) | Size (NoM) | Service Coupling (SC) | Testability |
|------|-----------------|------------|------------------------|-------------|
| RT1  | MED  | HIGH | LOW  | LOW |
| RT2  | MED  | HIGH | MED  | LOW |
| RT3  | MED  | HIGH | HIGH | LOW |
| RT4  | HIGH | LOW  | HIGH | LOW |
| RT5  | HIGH | MED  | LOW  | LOW |
| RT6  | HIGH | MED  | MED  | LOW |
| RT7  | HIGH | MED  | HIGH | LOW |
| RT8  | HIGH | HIGH | LOW  | LOW |
| RT9  | HIGH | HIGH | MED  | LOW |
| RT10 | HIGH | HIGH | HIGH | LOW |
| RT11 | LOW  | LOW  | HIGH | MED |
| RT12 | LOW  | MED  | HIGH | MED |
| RT13 | LOW  | HIGH | LOW  | MED |
| RT14 | LOW  | HIGH | MED  | MED |
| RT15 | LOW  | HIGH | HIGH | MED |
| RT16 | MED | LOW | HIGH | MED |
| RT17 | MED | MED | LOW | MED |
| RT18 | MED | MED | MED | MED |
| RT19 | MED | MED | HIGH | MED |
| RT20 | HIGH | LOW | LOW | MED |
| RT21 | HIGH | LOW | MED | MED |
| RT22 | LOW | LOW | LOW | HIGH |
| RT23 | LOW | LOW | MED | HIGH |
| RT24 | LOW | MED | LOW | HIGH |
| RT25 | LOW | MED | MED | HIGH |
| RT26 | MED | LOW | LOW | HIGH |
| RT27 | MED | LOW | MED | HIGH |

b. Testability

Table 3. Fuzzy inference rules

After the fuzzification process, a certain metric value can simultaneously belong to multiple fuzzy sets. For example, in our model, based on the fuzzification parameters, the propagation cost of a microservice with a PC value of 21.8 belongs to fuzzy sets LOW and MED simultaneously because $\mu_L^{PC}(21.8) = 0.5$, $\mu_M^{PC}(21.8) = 0.5$, and $\mu_H^{PC}(21.8) = 0$. Therefore, it can fit all rules for testability where the service coupling input is either LOW or MED. Since metrics can belong to multiple fuzzy sets simultaneously, a microservice can satisfy multiple inference rules with different output values. To combine these rules, we use Mamdani's max-min inference method [58], which consists of two main steps, i.e., implication and aggregation. In the first step, the strength of each satisfied rule is calculated, taking the minimum of the membership value of the inputs. The equation used to calculate the strengths of the modifiability rules is given below in Equation 5.

$$S_{RMx} = min\left(\mu_A^{PC}(PC_m), \mu_B^{AC}(AC_m)\right) \quad (5)$$
$$A, B = \{L, M, or\ H\}$$

$S_{RMx}$: The strength of the inference rule RMx for modifiability
$PC_m$: The value of the metric PC for the microservice m
$AC_m$: The value of the metric AC for the microservice m

For example, if the metric values PC = x and AC = y of a particular microservice yield membership values $\mu_M^{PC}(x) = 0.2$ (0.2 MED) and $\mu_H^{AC}(y) = 0.4$ (0.4 HIGH), respectively, they meet the rule RM1 with the output LOW. The strength of this output is $S_{RM1}$ = min(0.2, 0.4) = 0.2. Since PC = x and AC = y values correspond to different fuzzy values simultaneously, they will satisfy multiple rules with different outputs and strengths. For example, the same metric value PC = x can also generate another membership value such as $\mu_L^{PC}(x) = 0.8$ (0.8 LOW). Now, PC and AC metrics will meet the rule RM4 with the output MED. The strength of this output is $S_{RM4}$ = min(0.8, 0.4) = 0.4. In the second step of Mamdani's method, these outputs are aggregated by taking the maximum strength values for each output set, as shown below.

$$\mu^{Mod} = \max_x(S_{RMx}), \quad For\ each\ active\ rule\ RMx \quad (6)$$

Here, $\mu^{Mod}$ denotes the aggregated fuzzy set for modifiability.

Similarly, the following equation is used to calculate the strengths of the testability rules.

$$S_{RTx} = min(\mu_A^{AC}(AC_m), \mu_B^{NoM}(NoM_m), \mu_C^{SC}(SC_m),) \quad (7)$$
$$A, B, C = \{L, M, or\ H\}$$

$S_{RTx}$: The strength of the inference rule RTx
$AC_m$: The value of the metric AC for the microservice m
$NoM_m$: The value of the metric NoM for the microservice m
$SC_m$: The value of the metric SC for the microservice m

The outputs of the satisfied testability rules are aggregated by taking the maximum strength values for each output set, as shown below:

$$\mu^{Test} = \max_x(S_{RTx}), \quad For\ each\ active\ rule\ RTx \quad (8)$$

where $\mu^{Test}$ denotes the aggregated fuzzy set for testability.

### 4.2.3. Defuzzification

The defuzzification process involves combining the aggregated fuzzy set obtained from FIM with a mathematical function, called a defuzzification method, to produce a single crisp value, which is a numerical value that can be used in calculations or decision-making. Membership functions are also needed in the defuzzification process to convert fuzzy values (LOW, MED, and HIGH) in the aggregated fuzzy set to numerical crisp values. In this study, we use the membership functions presented in Table 4 for both modifiability and testability. The shapes of the functions are illustrated in Figure 5.

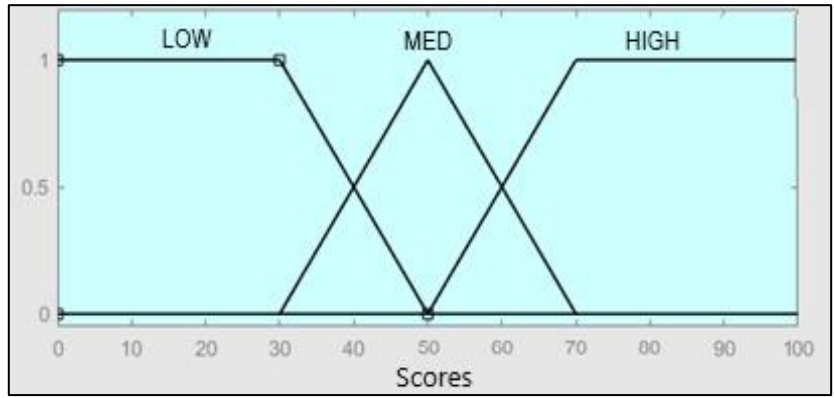

**Fig 5.** Membership functions for defuzzification

| Linguistic Variable | Membership Function | Parameters (Intervals) |
|---|---|---|
| LOW | Trapezoidal | [0 ,0, 30, 40] |
| MED | Triangular | [30, 40, 50] |
| HIGH | Trapezoidal | [50, 70, 100, 100] |

**Table 4.** Parameters of membership functions for defuzzification

We determined the functions and parameters given in Table 4 to make our method generate the scores presented in Table 5 depending on the level of the measured sub-characteristics modifiability and testability.

| Score | Meaning |
|---|---|
| 0 - 30 | LOW |
| 30 - 40 | LOW – MED, closer to LOW |
| 50 | MED – MED |
| 50 – 60 | HIGH closer to MED |
| 60 – 70 | MED – HIGH closer to HIGH |
| 70 – 100 | HIGH |

**Table 5.** Meanings of the scores generated by the method

We use these scores to interpret and evaluate the quality levels of the microservices. It is also possible to employ different parameters for the membership functions. Using different parameters will only change the generated scores and their interpretations.

To calculate the crisp values, we employ the centroid method, one of the most commonly used defuzzification methods [59]. In the centroid method, the crisp value is calculated as the center of gravity of the aggregated fuzzy set, representing the overall "weight". Based on the parameters we selected for defuzzification, the centroid method generates scores between 17.4 and 82.6.

Figure 6 presents the entire fuzzy measurement process for the modifiability level of an exemplary microservice. In this example, a specific microservice under evaluation has the metric values PC = 23.4 and AC = 6.53. After the fuzzification process, the membership degrees of PC to fuzzy sets are as follows: LOW: 0, MED: 0.8, and HIGH: 0.2. As a result, we obtain QME1={0, 0.8, 0.2}. The given value of AC belongs only to set HIGH because its membership degree is 1 to HIGH and zero to LOW and MED. As a result, QME2 = {0, 0, 1}. These input values satisfy two fuzzy inference rules of modifiability, i.e., RM6 and RM9. According to Mamdani's max-min inference method, we calculate the strength of the RM6 using Equation 5 and obtain 0.8 in the implication (min) step [58]. Hence, the output of the rule RM6 is 0.8 LOW. Similarly, the strength of rule RM9 is calculated as 0.2 LOW. In the second step (max), we aggregate these two outputs using Equation 6 and obtain 0.8 LOW. In the defuzzification process, the centroid method applied to the aggregated fuzzy set yields a score of QM1=21.7 for the modifiability level of the microservice under evaluation. As this result is below a predetermined threshold of 40 (LOW-MED), we can conclude that the exemplary microservice evaluated in Figure 6 is poorly designed for modifiability.

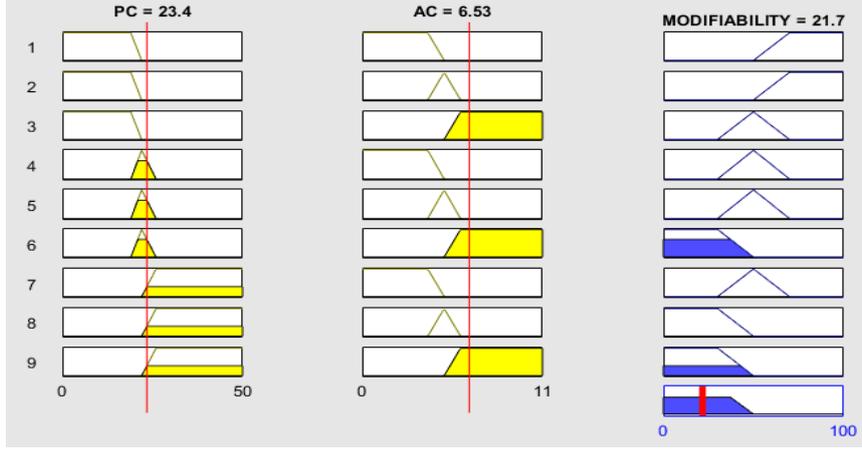

**Fig 6.** Fuzzy measurement process for the modifiability level of an exemplary microservice

### 4.3. Calculation of the Maintainability

After obtaining the scores of the sub-characteristics modifiability and testability, we calculate their weighted average using Equation 9 to determine the maintainability level of the microservices in SUE.

$$MNT_m = w_{mod} \times MOD_m + w_{tst} \times TST_m \qquad (9)$$

The meanings of the symbols in Equation 9 are given below.
$MNT_m$ represents the overall maintainability score of a microservice m.
$MOD_m$ denotes the score of the modifiability of the microservice m, i.e., the quality measure QM1 obtained during the defuzzification.
$TST_m$ signifies the score of the testability of the microservice m, i.e., the quality measure QM2 obtained during the defuzzification.

The parameters $w_{mod}$ and $w_{tst}$ are the weights assigned to modifiability and testability. These weights lie in the range [0-1], and their sum is always equal to 1, ensuring that the combined score is a normalized value. In our experiments, we set both wmod and wtst to 0.5, placing equal emphasis on modifiability and testability in microservice maintainability score computation. Companies utilizing our method can use different weights depending on the importance they attach to modifiability and testability in their projects.

To decide if a microservice needs refactoring due to a low maintainability level, we compare its score to a predetermined threshold $T_{REF}$. When the maintainability score of microservices is smaller than or equal to the threshold, they should be considered for refactoring. This can be determined using the following expression:

If $MNT_m \leq T_{REF}$, the microservice m needs refactoring.

In our study, we selected $T_{REF}=40$ because the membership functions we use in the defuzzification process generate scores below 40 if the sub-characteristics tend to be LOW, as presented in Figure 5. However, companies can also use other threshold values, for example, 30 if they want to deal only with poorly designed microservices, or higher values, such as 50, if they also intend to improve microservices at a medium level of quality.

## 5. EXPERIMENTS AND EVALUATION

To evaluate our approach, we applied our model to an MSA-based open-source project, Train Ticket[3], developed in Java. We compared our results with the opinions of three software developers who analyzed the project and categorized the microservices in advance.

The Train Ticket application provides standard train ticket booking capabilities, such as ticket inquiry, reservation, payment, change, and user notification [60]. It is a well-known project and is used as a benchmark microservice system by many studies in the literature [17], [61-69]. The Train Ticket was specifically chosen as our test project due to its medium size and tailored design to represent a real-world microservice application. Another reason to choose this project is that TrainTicket was developed using the same technology as the reference projects. Although the system contains 41 microservices, we used 36 of them in our experiments. The other five do not align with the comprehensive criteria for defining microservices [70], so they were deemed potential microservice candidates and were excluded from this study.

### 5.1. Experimental Results

Table 6 presents the results of our model applied to the Train Ticket project. To make referencing the microservices easier, we assigned them numbers in the first column of the table. The second column contains the names of the microservices. The columns PC, AC, NoM, and SC present the corresponding metrics we collected from the microservices. The last three columns present the results yielded by our method. Modifiability and Testability columns contain the scores obtained in the defuzzification step for the corresponding sub-characteristics. Maintainability is the weighted average of the sub-characteristics. In this experiment, Modifiability and Testability have the same weights, i.e., $w_{mod} = w_{tst} = 0.5$. Since we selected $T_{REF}=40$, maintainability scores smaller than or equal to 40 are bolded. These results indicate that the following nine microservices have maintainability problems and should be refactored: M9, M14, M15, M16, M21, M22, M24, M27, and M36.

| MS. Num. | Microservice Name | PC | AC | NoM | SC | Modifiability | Testability | Maintainability |
|---|---|---|---|---|---|---|---|---|
| **M1** | AdminBasicInfoService | 27 | 1 | 67 | 0.58 | 50.00 | 50.00 | 50.00 |
| **M2** | AdminOrderService | 20.31 | 3 | 19 | 0.14 | 76.57 | 70.99 | 73.78 |
| **M3** | ts-admin-route-service | 26.53 | 1 | 17 | 0.11 | 57.06 | 80.27 | 68.66 |
| **M4** | ts-admin-travel-service | 21.53 | 2.1 | 23 | 0.14 | 77.69 | 67.75 | 72.72 |
| **M5** | ts-admin-user-service | 18 | 1 | 19 | 0.17 | 79.84 | 74.79 | 77.31 |
| **M6** | ts-assurance-service | 17.36 | 1.97 | 49 | 0.25 | 79.84 | 50.00 | 64.92 |
| **M7** | ts-auth-service | 13.75 | 4.5 | 44 | 0.25 | 79.25 | 21.97 | 50.61 |

[3] https://github.com/FudanSELab/train-ticket

| | | | | | | | | |
|---|---|---|---|---|---|---|---|---|
| M8 | ts-basic-service | 19.9 | 4.47 | 24 | 0.08 | 75.58 | 37.42 | 56.50 |
| M9 | ts-cancel-service | 10.73 | 6.27 | 25 | 0.08 | 50.00 | 18.18 | **34.09** |
| M10 | ts-config-service | 26.56 | 1.54 | 27 | 0.17 | 56.67 | 59.66 | 58.17 |
| M11 | ts-consign-price-service | 26.56 | 1.58 | 21 | 0.14 | 56.67 | 71.70 | 64.18 |
| M12 | ts-consign-service | 18 | 1.66 | 30 | 0.17 | 79.84 | 50.00 | 64.92 |
| M13 | ts-contacts-service | 23.46 | 1.52 | 41 | 0.22 | 79.39 | 50.00 | 64.70 |
| M14 | ts-execute-service | 23.44 | 6.53 | 17 | 0.08 | 22.81 | 17.52 | **20.17** |
| M15 | ts-food-delivery(map)-service | 27.08 | 4.4 | 34 | 0.19 | 24.72 | 25.85 | **25.29** |
| M16 | ts-food-service | 20.14 | 9.68 | 39 | 0.19 | 40.36 | 18.39 | **29.38** |
| M17 | ts-inside-payment-service | 12.81 | 3.98 | 69 | 0.25 | 78.09 | 29.67 | 53.88 |
| M18 | ts-notification-service | 19 | 1 | 37 | 0.14 | 79.84 | 50.00 | 64.92 |
| M19 | ts-order-other-service | 18.21 | 3.96 | 78 | 0.44 | 78.05 | 29.95 | 54.00 |
| M20 | ts-order-service | 18.21 | 3.96 | 84 | 0.44 | 78.05 | 29.95 | 54.00 |
| M21 | ts-preserve-other-service | 8.22 | 10.94 | 23 | 0.06 | 50.00 | 18.53 | **34.26** |
| M22 | ts-preserve-service | 7.91 | 10.99 | 23 | 0.06 | 50.00 | 18.53 | **34.26** |
| M23 | ts-price-service | 22.22 | 2.06 | 32 | 0.17 | 78.32 | 50.00 | 64.16 |
| M24 | ts-rebook-service | 11.11 | 7.37 | 28 | 0.08 | 50.00 | 17.69 | **33.84** |
| M25 | ts-route-plan-service | 16.89 | 3.17 | 20 | 0.11 | 79.08 | 65.26 | 72.17 |
| M26 | ts-route-service | 23.46 | 3.17 | 34 | 0.17 | 75.52 | 41.42 | 58.47 |
| M27 | ts-seat-service | 16.67 | 5.57 | 15 | 0.08 | 53.48 | 25.70 | **40.00** |
| M28 | ts-security-service | 16.67 | 1.64 | 30 | 0.17 | 79.84 | 50.00 | 64.92 |
| M29 | ts-station-service | 26.56 | 1.9 | 41 | 0.25 | 56.67 | 50.00 | 53.34 |
| M30 | ts-ticketinfo-service | 26.45 | 1 | 13 | 0.08 | 58.06 | 81.65 | 69.86 |
| M31 | ts-train-service | 26.56 | 1.47 | 36 | 0.17 | 56.67 | 50.00 | 53.34 |
| M32 | ts-travel-plan-service | 14.53 | 1.91 | 32 | 0.14 | 79.84 | 50.00 | 64.92 |
| M33 | ts-travel-service | 14.37 | 2.22 | 66 | 0.33 | 79.84 | 50.00 | 64.92 |
| M34 | ts-travel2-service | 14.37 | 2.62 | 60 | 0.33 | 79.84 | 50.00 | 64.92 |
| M35 | ts-user-service | 19.83 | 2.17 | 38 | 0.28 | 79.15 | 50.00 | 64.57 |
| M36 | ts-verification-code-service | 25 | 5.71 | 16 | 0.06 | 21.58 | 17.37 | **19.47** |

Table 6. Metric values and scores of microservices in the Train Ticket application

### 5.2. Validation of the Results

To validate the accuracy of our findings, we sought the assistance of three experienced software developers with over five years of expertise in implementing microservice-based systems. One of the developers also has a Ph.D. degree in the subject. They independently evaluated the maintainability of each microservice, considering their evaluation criteria, such as deployment structures, service interactions, and service implementations. The evaluators classified the microservices into three categories: LOW (L), MED (M), and HIGH (H). Microservices with major maintainability issues that should be refactored are categorized as LOW. Microservices in the MED category may have some minor maintainability issues but may also be well-designed in certain aspects. These microservices do not need refactoring in the short term. The HIGH category is for well-designed microservices without any maintainability problems. Since the evaluators classified microservices independently, some microservices were labeled differently. However, the labels were generally consistent, and at least two evaluators always reached the same conclusion for a particular microservice. We reached a final decision by forming a consensus based on the majority opinion of the evaluators.

Table 7 presents the labels assigned to the microservices by human evaluators and the maintainability scores calculated by the proposed method. The first column contains the same number of microservices as in Table 6. The columns E1, E2, and E3 present the labels assigned by the human evaluators. "Decision" is the final decision based on the majority opinion of the evaluators. The last column contains the maintainability scores of the microservices calculated using the proposed method.

| MS. Num. | E1 | E2 | E3 | Decision | Maintainability by Model | MS. Num. | E1 | E2 | E3 | Decision | Maintainability by Model |
|---|---|---|---|---|---|---|---|---|---|---|---|
| M1 | M | H | M | M | 50.00 | M19 | M | M | M | M | 54.00 |
| M2 | H | H | H | H | 73.78 | M20 | M | L | M | M | 54.00 |
| M3 | M | H | H | H | 68.66 | M21 | L | L | M | L | **34.26** |
| M4 | H | H | H | H | 72.72 | M22 | L | L | M | L | **34.26** |
| M5 | H | H | H | H | 77.31 | M23 | M | H | H | H | 64.16 |
| M6 | M | M | M | M | 64.92 | M24 | L | L | L | L | **33.84** |
| M7 | M | H | M | M | 50.61 | M25 | H | M | H | H | 72.17 |
| M8 | M | L | M | M | 56.50 | M26 | H | H | M | H | 58.47 |
| M9 | M | L | L | L | **34.09** | M27 | M | L | M | M | **39.59** |
| M10 | H | H | H | H | 58.17 | M28 | H | H | H | H | 64.92 |
| M11 | M | H | H | H | 64.18 | M29 | H | H | H | H | 53.34 |
| M12 | M | H | H | H | 64.92 | M30 | H | H | H | H | 69.86 |
| M13 | H | H | M | H | 64.70 | M31 | H | H | M | H | 53.34 |
| M14 | M | M | L | M | **20.17** | M32 | H | M | H | H | 64.92 |
| M15 | L | L | L | L | **25.29** | M33 | M | L | M | M | 64.92 |
| M16 | L | L | M | L | **29.38** | M34 | M | M | M | M | 64.92 |
| M17 | M | M | M | M | 53.88 | M35 | M | H | M | M | 64.57 |
| M18 | H | H | M | H | 64.92 | M36 | L | L | M | L | **19.47** |

**Table 7.** Comparison of the quality model's results with the assessment of the evaluators

As the main objective of this study is to identify microservices that must be refactored due to maintainability issues, we first compared nine microservices with a maintainability score of less than or equal to 40 with those categorized as LOW by the evaluators. According to the "Decision" column of Table 7, the evaluators classified the following seven microservices as LOW: M9, M15, M16, M21, M22, M24, and M36. On the other hand, the proposed method detected the following nine microservices: M9, M14, M15, M16, M21, M22, M24, M27, and M36.

Notably, our model successfully identified all seven instances of low-maintainable microservices. However, it classified two microservices, M14 and M27, as LOW, while the developers considered them to be MED. The confusion matrix [71] in Table 8 presents the performance of the method in detecting microservices with a low maintainability level.

|  | | Evaluators | |
|---|---|---|---|
|  | | LOW | Not LOW |
| **Predicted (Model)** | LOW | TP = 7 | FP = 2 |
|  | Not LOW | FN = 0 | TN = 27 |
|  | Total | 7 | 29 |

Table 8. Confusion matrix

The meanings of entries in the confusion matrix are:
- TP (True Positive) is the number of microservices labeled as LOW by the evaluators and predicted correctly by the method.
- FP (False Positive) is the number of microservices NOT labeled as LOW by the evaluators and predicted incorrectly by the method.
- FN (False Negative) is the number of microservices labeled as LOW by the evaluators and predicted incorrectly by the method.
- TN (True Negative) is the number of microservices NOT labeled as LOW by the evaluators and predicted correctly by the method.

To evaluate the performance of our prediction models, we computed recall, precision, F-Measure, and accuracy, which are defined in the following equations using the values from the confusion matrix in Table 8.

Recall = TP / (TP + FN)
Precision = TP / (TP + FP)
F-Measure = 2 * Precision * Recall / (Precision + Recall)
Accuracy = (TP + TN) /( TP + TN + FP + FN)

| Recall | Precision | F-Measure | Accuracy |
|---|---|---|---|
| 100% | 77.78% | 87.5% | 94.44% |

Table 9. Performance of the method in detecting microservices with low maintainability level on the Train Ticket project

The results in Table 9 obtained from the experiment on the Train Ticket project show that the proposed method can be trusted to guide developers in detecting microservices with a low maintainability level.

In addition, we examined the maintainability scores of the microservices classified as HIGH and MED and found that they were consistent with the opinion of the evaluators. The averages and standard deviations of the maintainability scores calculated by our method for the microservices with the particular labels assigned by the evaluators are given in Table 10.

| Label by the Evaluators | Average | Standard Deviation |
|---|---|---|
| HIGH | 65.35 | 6.64 |
| MED | 53.22 | 12.39 |
| LOW | 29.87 | 5.52 |

**Table 10.** The averages and standard deviations of the maintainability scores for the microservices with particular labels

Table 10 indicates that the method's maintainability scores align with the targeted meanings in Table 5. In line with the study's main purpose, scores assigned to low-labeled microservices correspond to areas classified as "low" or "low-medium, closer to low".

The findings indicate that the proposed method can assign maintainability scores to microservices that correlate with human evaluators' opinions and detect microservices requiring refactoring.

## 6. DISCUSSION

The obtained results show that our fuzzy logic model successfully detected all seven microservices labeled as LOW, achieving a perfect recall score of 100%. On the other hand, two microservices were classified as "LOW" by our model but considered "MED" by developers, leading to 77.78% precision, which resulted in 94.44% accuracy in detecting the microservices labeled as LOW. We analyzed the two microservices, M14 and M27, that our method categorized as LOW, while the developers considered them to be MED. We observed that one of the three developers categorized both of them as LOW as well. The computed score of M27 is 39.59, which is around the threshold between LOW and MED. Given these results, the model successfully identifies all microservices labeled as LOW. This success aligns with our primary objective of predicting low-labeled microservices, highlighting the effectiveness of our approach in addressing maintainability concerns.

The results achieved in the Train-ticket project could not be directly compared with existing literature due to the lack of studies focusing on individual microservice evaluation. However, to illustrate the benefits of a complete fuzzy logic system for the measurement of maintainability, we also conducted measurements on the same case study using classical logic. The classical logic contains a basic categorization process, where a crisp metric value can belong only to one set. Similarly to the original fuzzification process, we use three distinct quartiles, Q1, Q2, and Q3, to obtain the categorization parameters for the classical logic. For each metric array, the average of the first quartile (Q1) and second quartile (Q2) corresponds to the border between LOW and MED. The average of the second quartile(Q2) and third quartile

(Q3) represents the border between MED and HIGH. To emphasize the differences between classical logic and our approach, we included Figure 7, showing the categorization parameters (above) alongside the reintroduced general shapes of membership functions of the original fuzzification (below).

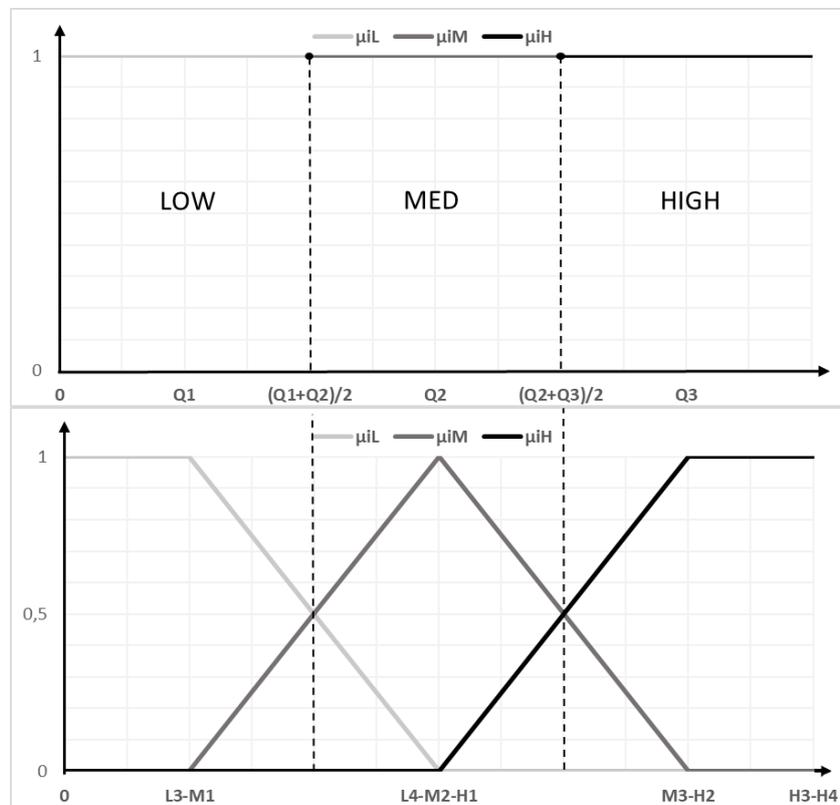

**Figure 7.** Categorization parameters of classical logic (above) and fuzzification parameters of the proposed method (below).

In the evaluation conducted with classical logic, microservices MS7 and MS8, labeled as MED by consensus and confirmed as such by our model, were incorrectly classified as LOW. This resulted in a total of 4 false positives, with a precision of 63.64% and an accuracy of 88.89% for classical logic. In contrast, when assessed using fuzzy logic, the precision of our model improved to 77.78%, with an accuracy of 94%. Although both models achieved a recall value of 100%, fuzzy logic outperformed classical logic by identifying fewer false positives and attaining a higher precision. Given the study's focus on identifying low-level microservices, fuzzy logic was chosen over classical logic, acknowledging the precision-recall tradeoff. The flexibility of fuzzy logic, which allows metrics to belong to multiple fuzzy set ranges, enabled our model to accurately detect microservices requiring refactoring while minimizing false positives. Consequently, the utilization of fuzzy logic resulted in a significant increase in precision from 63.64% to 77.78%, thereby enhancing the overall performance and accuracy of the model.

In conclusion, our findings indicate the potential application of fuzzy logic-based measurement systems to various aspects of microservice quality assessment using proper quality models. The success of our model in a widely referenced application like Train Ticket illustrates its practical utility in predicting the maintainability of individual microservices during the development of the software system.

## 7. THREATS TO VALIDITY

In this section, we highlight certain threats to the validity of our measurement system, which could potentially influence the results of the proposed quality assessment model.

**Determining the parameters for fuzzification from reference projects automatically:**

The first threat to the validity of the results comes from the process of determining the parameters for the fuzzification of metrics. To define the parameters of the membership functions, $\mu_L^i(x)$, $\mu_M^i(x)$, and $\mu_H^i(x)$ automatically, we use statical data obtained from 14 microservices of two MSA-based open-source reference projects developed in Java, i.e., "Tea Store" and "Microservice Observability". The limited number of reference projects and microservices may pose a threat to the validity of the measurements, as more industrial and open-source projects are needed to capture the general characteristics of microservices. Nevertheless, we utilized these projects as examples, recognizing the need to increase their number for a more comprehensive analysis.

The reason these two projects were chosen as reference projects is that they included services of varying quality. For the evaluation system to function correctly, microservices in reference projects have different levels of quality characteristics. Thus, the microservices selected from reference projects were intentionally diverse, representing a spectrum of quality characteristics. The potential bias in selecting projects based on observed quality characteristics may impact external validity. If the reference projects predominantly represent either high or low-quality microservices, the proposed model's parameters may lack accuracy when applied to a more diverse set of projects. This dependency introduces potential variability and may hinder the model's applicability to different projects, emphasizing the need for a broader analysis of metrics and characteristics. If the reference projects predominantly exhibit high or low-quality microservices, internal validity is compromised. To achieve more realistic results, gathering data from a wider range of industrial and open-source projects would be beneficial.

In addition, when developing a measurement system to assess software systems for particular applications that include microservices with specific design aspects, it is recommended to select reference projects with similar features to collect the overall characteristics of the microservices.

**Exclusive reliance on Train Ticket and developers for evaluation:**

Using the Train Ticket project for evaluation poses a threat to validity, as results may not be applicable across different projects. Although the Train Ticket project is one of the largest available open-source microservice projects, there is still a validity threat that we cannot necessarily generalize the results we obtained in our experiments to other software projects. To ensure the generalizability of the results, testing a larger sample of projects with a more extensive set of microservices reflecting diverse characteristics would be beneficial. To mitigate these threats to validity, we plan to test our approach on a variety of different projects and contexts to ensure that it is applicable and useful beyond the initial projects that were used to develop and test the model.

Another threat to the validity may arise from the group of evaluators who manually categorized the microservices. To ensure the validation process was robust, we enlisted the help of three experienced software developers with more than five years of experience in developing microservice-based systems. However, there is still a possibility another group of developers could categorize the same microservices slightly differently. Despite these threats, the results of our approach can still be used to infer a direction for the development and evaluation of microservices.

**Technology Platform Dependency:**

Another threat to validity arises from the potential variations in implementation techniques across projects, influenced by factors such as technology platforms, frameworks, and system complexity. It is essential to acknowledge that microservices projects can be developed using different platforms and technologies, leading to unique characteristics [72]. To ensure a meaningful evaluation, we carefully chose projects with similar features for reference and testing. Given that Maven[4] is one of the most widely used platforms for developing microservices, we selected microservice projects built with Maven for our study. Since we tested our method only on projects developed using a single platform, it may raise concerns about generalizing the model's findings to different projects with distinct architectural features. However, our method allows users also to select reference and test projects from alternative platforms if desired. It is important to note here that reference and test projects must have similar features to ensure a meaningful evaluation.

The threats to validity mentioned above result from the restricted availability of microservice-based projects on publicly accessible platforms like GitHub. Despite the existence of many projects, a few microservice projects are suitable for evaluation. This is due to the fact that some projects are too small and don't operate independently, which is against the general characteristics of MSA. Moreover, to utilize the SonarGraph tool, it is essential to run the entire project, but some projects are not fully uploaded to GitHub and therefore are not suitable for selection. Addressing these threats to validity requires a more diverse dataset for parameter determination, testing on a larger range of projects, careful consideration of technology platform dependencies, and a comprehensive approach to identifying microservices characteristics. We acknowledge the importance of these considerations and plan to address them in future research and model refinement.

---

4. https://maven.apache.org/

## 8. CONCLUSION

In this study, we proposed a novel approach to measure and evaluate the maintainability of MSAs employing fuzzy logic principles. Our method offers several distinct advantages. Firstly, it provides a comprehensive and refined assessment of MSA maintainability by leveraging low-level quantifiable software properties. We followed the ISO/IEC 25010 and 25020 standards to create a hierarchical model that defines the relations between the high-level characteristic maintainability, the sub-characteristics modifiability and testability, and related software metrics. Furthermore, the proposed method addresses the inherent ambiguity in interpreting these metrics by employing fuzzy logic techniques, i.e., fuzzification, inference, and defuzzification. We evaluated the effectiveness of our method by applying it to an open-source MSA-based project, Train Ticket. We obtained promising results that align closely with manual assessments conducted by experienced developers. These results indicate that our proposed model is a useful tool for development teams, as it helps identify low maintainability and facilitates refactoring strategies, contributing to overall software quality improvement.

A noteworthy direction for future research would be to empirically assess the model's effectiveness on a broader range of industrial and open-source projects. Such validations could provide critical insights into how the fuzzification membership functions can be refined to improve evaluation accuracy. Additionally, the fuzzy inference rules used in the model could be adjusted to improve the evaluation of quality characteristics and sub-characteristics. These enhancements would allow our model to generate more precise evaluations, ultimately contributing to the development of higher-quality MSAs. Our proposed hierarchical quality model holds promise for assessing the maintainability of MSAs and guiding future research toward improving the quality of microservice architectures.